\documentclass[aps,prd,onecolumn,groupedaddress,nofootinbib,showpacs,showkeys,amssymb]{revtex4}
\newlength{\extraspace}
\newlength{\extraspaces}
\begingroup
\endgroup

\newcommand{\be}{\begin{equation}
\addtolength{\abovedisplayskip}{\extraspaces}
\addtolength{\belowdisplayskip}{\extraspaces}
\addtolength{\abovedisplayshortskip}{\extraspace}
\addtolength{\belowdisplayshortskip}{\extraspace}}
\newcommand{\ee}{\end{equation}}

\newcommand{\ba}{\begin{eqnarray}
\addtolength{\abovedisplayskip}{\extraspaces}
\addtolength{\belowdisplayskip}{\extraspaces}
\addtolength{\abovedisplayshortskip}{\extraspace}
\addtolength{\belowdisplayshortskip}{\extraspace}}
\newcommand{\ea}{\end{eqnarray}}

\newcommand{\nonu}{\nonumber \\[.5mm]}
\newcommand{\A}{&\!\!\!}

\begin{document}

\title{ $D$-dimensional charged Anti-de-Sitter  black holes in  $f(T)$ gravity}
\author{  A. M. Awad$^{1,2}$, S. Capozziello$^{4,5,6}$ and  G.  G. L. Nashed$^{1,3}$}

\affiliation{ $^1$Center for Theoretical Physics, British University of Egypt,
Sherouk City 11837, Egypt\\
$^2$Department of Physics, Faculty of Science, Ain Shams University, Cairo 11566, Egypt\\
$^3$Department of Mathematics, Faculty of Science, Ain Shams University, Cairo 11566, Egypt\\
$^4$Dipartimento di Fisica ``E. Pancini``, Universit\'a di Napoli ``Federico II'',
Complesso Universitario di Monte Sant' Angelo, Edificio G, Via Cinthia, I-80126, Napoli, Italy\\
$^5$ Istituto Nazionale di Fisica Nucleare (INFN),  Sezione di Napoli,
Complesso Universitario di Monte Sant'Angelo, Edificio G, Via Cinthia, I-80126, Napoli, Italy\\
$^6$ Gran Sasso Science Institute, Viale F. Crispi, 7, I-67100, L'Aquila, Italy.\\
}

\begin{abstract}
 We present a $D$-dimensional  charged Anti-de-Sitter black hole solutions  in $f(T)$ gravity, where $f(T)=T+\beta T^2$ and $D \geq 4$. These  solutions are characterized by flat or cylindrical horizons. The interesting feature of these solutions is the existence of inseparable electric monopole and quadrupole terms in the potential which share related momenta, in contrast with most of the known charged black hole solutions in General Relativity and its extensions. Furthermore, these solutions have curvature singularities which are milder than those of the known  charged black hole solutions in General Relativity and Teleparallel  Gravity. This feature can be  shown by calculating some invariants of curvature and torsion tensors. Furthermore, we calculate the total energy of these black holes using the energy-momentum tensor. Finally, we show that these charged black hole solutions violate the first law of thermodynamics in agreement with previous results.
\end{abstract}

\maketitle

\section{Introduction}\label{S1}
 In the last decades, various investigations have shown that black hole physics plays a vital role in  our understanding of gravity on both macroscopic and microscopic scales. Aspects of black holes such as horizon's global structure, Hawking radiation, thermodynamical properties and black hole information are key concepts related to the fundamental structure of space-time. For recent review see Ref. (\cite{carlip} and references therein). Examples of these new concepts and constructions are the Holographic Principle and its realization in AdS/CFT correspondence \cite{thooft,susskind,maldacena}, Ashtekar approach and loop quantum gravity \cite{rovelli}, Jacobson's thermodynamic formulation \cite{jacobson} as well as Verlinde's ENTOPIC approach to gravity \cite{verlinde}. In general, seeking for new black hole solutions  is extremely relevant to set up  any relativistic theory of gravity.
 Recently, a lot of attention has been devoted to the  teleparallel equivalent of General Relativity (TEGR) \cite{Ea,Ea1,Ea2,HS9}. This theory is an equivalent  to General Relativity (GR) and can be generalized into a class of theories called $f(T)$ gravity \cite{FF7,FF8,BF9,CCMS}. TEGR and its generalizations are based on Weitzenb\"{o}ck connection instead of Levi-Civita connection and use a vielbein field, ${e^a}_{\mu}$, as a fundamental field variable, instead of the metric \cite{N10}. TEGR is equivalent to GR since its field equations as well as its test particle equation of motion are equivalent to that of GR. In TEGR and its generalizations the gravitational field is represented by torsion instead of curvature because the later vanishes in these theories. TEGR can be constructed as a gauge theory of the translation group, where the vielbein is the gauge field and the torsion is the field strength \cite{APVB,AP}. This theory is invariant under diffeomorphism and local Lorentz transformations \cite{M1,Mj}.
 
There is a considerable interest in generalizing TEGR  in cosmological  through adding higher-torsion terms. These terms can mimic a  dark energy fluid in Friedman-Robertson-Walker (FRW) cosmologies without introducing  exotic equations of state \cite{bamba,CC23,CC,FF11,Ngrg02,GSV,WNE1,RHTMM,GG2,JRH,GG3,CC8,CC9,CC10,AN,FLR,BFG,NNC,Km,KS,BMT,DWM}. Teleparallel gravity with higher-torsion terms  depending only on the torsion scalar, $T$, are known as $f(T)$ theories. A part from these cosmological applications some interesting black holes solutions were discovered and studied in the context of $f(T)$ theories. One of the most studied models of these theories is $f(T)=T+\beta \,T^2$, where spherically symmetric approximate solution was found and analyzed \cite{SOE,Ncpl,IS,IS1,XD}, furthermore, the higher-torsion correction term was constrained in this theory using the Solar System data.

Exact solutions in higher curvature/ torsion theories are not always easy to find also in the framework of $f(T)$  \cite{Nprd3,CGSV13, Ngrg3}. Black holes with cosmological constant present some attractive features: for example various horizon topologies appear in contrast to the asymptotically flat case. In these cases, black hole horizons  can be spherical, hyperbolic or planar: these features can  lead to tori or cylindrical structures depending on the global identifications applied \cite{Mrb}. Asymptotically de Sitter and Anti-de-Sitter charged black holes and rotating black holes in four and higher dimensions have been intensively studied in the context of AdS/CFT and dS/CFT correspondences; see for example Refs. \cite{lemos,Aa02,Aa06,Aa07,Aa08,gubser}.

In this paper, we are going to  present  new charged asymptotically Anti-de-Sitter black hole solutions with flat horizons in $f(T)$ theories, where $f(T)=T+\beta T^2$,  $\beta <0$,  and $D \geq 4$. First of all, these black holes solutions  have electric potentials which contains a monopole  as well as a quadrupole term. In order to get an asymptotically Anti-de-Sitter (AdS) solution, we are forced to relate the electric monopole momentum and the quadrupole momentum of these solutions, therefore, these two momenta are inseparable. The second interesting feature is that, although this black hole solution is singular at the origin, $r=0$, its singularity is clearly milder than asymptotically Anti-de-Sitter charged black solution in GR or TEGR. For example in {\it D} dimensions, the Kretschmann scalar, derived from the Ricci tensor square,  and the Ricci  scalar  are $K=R_{\mu \nu}R^{\mu \nu} \sim r^{-2(D-2)}$, $R\sim  r^{-(D-2)}$  in contrast with, the known solutions in Einstein-Maxwell theory in both GR and TEGR which have $K =R_{\mu \nu}R^{\mu \nu}\sim r^{-2D}$ and $R \sim  r^{-D}$. Furthermore, although these asymptotically AdS charged solutions have different $g_{tt}$ and $g^{rr}$ component for the metric, they have coinciding Killing and event horizons.

 The outline of the paper  is  the following. In Section \ref{S1.1}, we briefly review the TEGR formalism through tensors definitions and the field equations, then we introduce the field equations of $f(T)$  gravity. In  Section \ref{S3}, a  vielbein field having flat horizon in $D$ dimensions is applied to the  field equations of $f(T)$ gravity to obtain a general neutral black hole solution in {\it D} dimensions, which is asymptotically AdS.  In Section \ref{S9}, a cylindrically symmetric vielbein  is applied to the Einstein-Maxwell field equations in $f(T)$ gravity.  We show how this solution can be reduced to an exact charged static black hole in AdS space. The interesting feature of this black hole is that it has  monopole  and the quadrupole momenta. This feature extends results presented in \cite{CGSV13}. In Section \ref{S5}, some relevant physical features of these black holes are reported. In  Section \ref{S11}, we discuss the thermodynamics of the black holes presented in \ref{S9}. Finally, concluding remarks are reported in Section \ref{S12}.

\section{The TEGR geometry and $f(T)$ gravity}\label{S1.1}
TEGR  is described by  the pair $\{{\cal M},~e_{i}\}$, with $\cal M$ being a $D$-dimensional manifold and $e_{i}$ ($i=1,2,\cdots, D$) are vectors globally defined  on the manifold $\cal M$. Vector fields  $e_{i}$ are the parallel  vectors. In $D$-dimension, the parallel  vectors  are named the {\it vielbein} fields and the derivative of the  contravariant vielbein fields have to vanish
\begin{equation}\label{q1}
  D_{\mu} {e_i}^\nu:=\partial_{\mu}
{e_i}^\nu+{\Gamma^\nu}_{\lambda \mu} {e_i}^\lambda= 0,
\end{equation}
where the  differentiation is with respect to the Weitzenb\"{o}ck connection and  $\partial_{\mu}:=\frac{\partial}{\partial x^{\mu}}$ and ${\Gamma^\nu}_{\lambda \mu}$ is a non-symmetric affine connection defined as \cite{Wr}
\begin{equation}\label{q2}
{\it {\Gamma^\lambda}_{\mu \nu} := {e_i}^\lambda~ \partial_\nu e^{i}{_{\mu}}}.
\end{equation}
In this geometry,  the metric is given by
\begin{equation}\label{q3}
 {\it g_{\mu \nu} :=  \eta_{i j} {e^i}_\mu {e^j}_\nu,}
\end{equation}
where $\eta_{i j}=(+,-,-,- \cdots)$ is  {\it D} dimensional Minkowskian metric. The metricity condition is fulfilled as a consequence of Eq. (\ref{q1}). The torsion, ${T^\alpha}_{\mu \nu}$, and the contortion, $K^{\mu \nu}{}_\alpha$, tensors field  are defined as
\begin{eqnarray}
\nonumber {T^\alpha}_{\mu \nu}  & := &
{\Gamma^\alpha}_{\nu \mu}-{\Gamma^\alpha}_{\mu \nu} ={e_i}^\alpha
\left(\partial_\mu{e^i}_\nu-\partial_\nu{e^i}_\mu\right),\\
{K^{\mu \nu}}_\alpha  & := &
-\frac{1}{2}\left({T^{\mu \nu}}_\alpha-{T^{\nu
\mu}}_\alpha-{T_\alpha}^{\mu \nu}\right). \label{q4}
\end{eqnarray}
We introduce the teleparallel torsion scalar of TEGR theory which is
\begin{equation}\label{Tor_sc}
T := {T^\alpha}_{\mu \nu} {S_\alpha}^{\mu \nu},
\end{equation}
where the skew symmetric tensor ${S_\alpha}^{\mu \nu}$  is defined as
\begin{equation}\label{q5}
{S_\alpha}^{\mu \nu} := \frac{1}{2}\left({K^{\mu\nu}}_\alpha+\delta^\mu_\alpha{T^{\beta
\nu}}_\beta-\delta^\nu_\alpha{T^{\beta \mu}}_\beta\right).
\end{equation}
Using Eq. (\ref{q4}) it is possible to re-express  Eq. (\ref{q2}) as
\begin{equation}\label{contortion}
    {\Gamma^\mu}_{\nu \rho }=\left \{_{\nu  \rho}^\mu\right\}+{K^{\mu}}_{\nu \rho},
\end{equation}
 where the first term  is the Levi-Civita affine connection of  GR   while  the second one is derived from the contortion.

It is natural to extend TEGR theory  including higher torsion terms defining  a Lagrangian  $f(T)$ where $f$ is a function of the torsion invariant $T$:
\begin{equation}\label{q7}
{\cal L}=\frac{1}{2\kappa}\int |e|(f(T)-2\Lambda)~d^{D}x+\int |e|{\cal L}_{ em}~d^{D}x,
\end{equation}
where $\kappa$  is  a dimensional constant defined as $\kappa =2(D-3)\Omega_{D-1} G_D$, with  $G_D$  being  the Newton gravitational  constant in $D$-dimensions and
$\Omega_{D-1}$  the volume of  $(D-1)$-dimensional unit sphere given by
the expression $\Omega_{D-1} = \frac{2\pi^{(D-1)/2}}{\Gamma((D-1)/2)}$, with the $\Gamma$-function
of the argument that depends on the dimension of spacetime\footnote{When $D = 4$, one can show that $2(D-3)\Omega_{D-1} = 8 \pi$.}.  In Eq. (\ref{q7}), $ |e|=\sqrt{-g}=\det\left({e^a}_\mu\right)$ and ${\cal L}_{
em}=-\frac{1}{2}{ F}\wedge ^{\star}{F}$ is the Maxwell Lagrangian,
with  $F = dA$, with $A=A_{\mu}dx^\mu$, is the electromagnetic
potential 1-form \cite{CGSV13}. The variation of Eq. (\ref{q7}) with respect to the vielbein field ${e^i}_\mu$ and the vector potential $A_{\mu}$ gives the following field equations \cite{BF9}
\begin{eqnarray}\label{q8}
& &{S_\mu}^{\rho \nu} \partial_{\rho} T f_{TT}+\left[e^{-1}{e^i}_\mu\partial_\rho\left(e{e_i}^\alpha
{S_\alpha}^{\rho \nu}\right)-{T^\alpha}_{\lambda \mu}{S_\alpha}^{\nu \lambda}\right]f_T
-\frac{f-2\Lambda}{4}\delta^\nu_\mu +\frac{1}{2}\kappa{{{\cal
T}^{{}^{{}^{^{}{\!\!\!\!\scriptstyle{em}}}}}}}^\nu_\mu=H^\nu{}_\mu \equiv0,\nonumber\\
&&\partial_\nu \left( \sqrt{-g} F^{\mu \nu} \right)=0\; ,
\end{eqnarray}
where $f := f(T)$, \ \   $f_{T}:=\frac{\partial f(T)}{\partial T}$, \ \  $f_{TT}:=\frac{\partial^2 f(T)}{\partial T^2}$ and ${{{\cal
T}^{{}^{{}^{^{}{\!\!\!\!\scriptstyle{em}}}}}}}^\nu_\mu$ is the
energy-momentum tensor of the  electromagnetic field defined as
\[
{{{\cal
T}^{{}^{{}^{^{}{\!\!\!\!\scriptstyle{em}}}}}}}^\nu_\mu=F_{\mu \alpha}F^{\nu \alpha}-\frac{1}{4} \delta_\mu{}^\nu F_{\alpha \beta}F^{\alpha \beta}.\] Eq.  (9) can be re-expressed  as
\be \partial_\nu \Biggl[e{S}^{a \rho \nu} f_T\Biggr]=\kappa e
{e^a}_\mu \Biggl[t^{\rho \mu}+{{{\cal
T}^{{}^{{}^{^{}{\!\!\!\!\scriptstyle{em}}}}}}}^{\rho \mu}\Biggr],\ee
where $t^{\nu \mu}$ has the form \be t^{\nu
\mu}=\frac{1}{\kappa}\Biggl[4f_T {S^\alpha}^{\nu
\lambda}{T_{\alpha \lambda}}^{\mu}-g^{\nu \mu} f\Biggr].\ee Since   ${S}^{a \nu \lambda}$ is a skew-symmetric tensor in the last pairs,   then \be
\partial_\mu \partial_\nu\left[e{S}^{a \mu \nu} f_T\right]=0, \quad
 \textrm{which \quad yields}
\quad \partial_\mu\left[e\left(t^{a \mu}+{{{\cal
T}^{{}^{{}^{^{}{\!\!\!\!\scriptstyle{em}}}}}}}^{a
\mu}\right)\right]=0. \ee Eq. (12) yields \be
\frac{d}{dt}\int_V d^{(D-1)}x \ e \ {e^a}_\mu \left(t^{0 \mu}+{{{\cal
T}^{{}^{{}^{^{}{\!\!\!\!\scriptstyle{em}}}}}}}^{0
\mu}\right)+ \oint_\Sigma \left[e \ {e^a}_\mu \ \left(t^{j
\mu}+{{{\cal T}^{{}^{{}^{^{}{\!\!\!\!\scriptstyle{em}}}}}}}^{j
\mu}\right)\right]=0.\ee Eq. (13)  is the  conservation law of  the
energy-momentum tensor ${{{\cal
T}^{{}^{{}^{^{}{\!\!\!\!\scriptstyle{em}}}}}}}^{\lambda
\mu}$  and   the quantity $t^{\lambda \mu}$. Thus,  we can consider  $t^{\lambda \mu}$
 to describe  the gravitational energy-momentum tensor  in
  $f(T)$  gravity \cite{US3}.  Therefore,
the   energy-momentum of $f(T)$  theory contained
in a (D-1)-dimensional volume $V$ takes the form \be P^a=\int_V d^{(D-1)}x
\ e \ {e^a}_\mu \left(t^{0 \mu}+{{{\cal
T}^{{}^{{}^{^{}{\!\!\!\!\scriptstyle{em}}}}}}}^{0
\mu}\right)=\frac{1}{\kappa}\int_V d^{(D-1)}x  \partial_\nu\left[e{S}^{a 0
\nu} f_T\right].\ee From  Eq. (14),  we can return to the  standard TEGR as soon as    $f(T)=T$ \cite{MDTC}. Eq. (13) represents the conserved four-momentum for any asymptotic flat solution: in this work we discuss a class of asymptotically AdS solution. Therefore, it is natural to calculate conserved quantities relative to a pure AdS space. Otherwise, the conserved quantities is plagued by infinities because of the asymptotic behavior of the asymptotically AdS solution. For example the total mass/energy of an AdS black hole measured by a stationary observer at infinity might be understood as the difference in energy between the AdS black hole solution and AdS space itself. Therefore, in calculating conserved quantities,  it is natural to subtract off the contributions coming from pure AdS space in the above conserved quantities of Eq. (13); this is why we have subscript "$r$" which stands for the regularized value of the physical quantity.
\[
\frac{d}{dt}\int_V d^{(D-1)}x \ e \ {e^a}_\mu \left(t_{r}^{0 \mu}+{{{\cal
T}_{r}^{{}^{{}^{^{}{\!\!\!\!\scriptstyle{em}}}}}}}^{0
\mu}\right)+ \oint_\Sigma \left[e \ {e^a}_\mu \ \left(t_{r}^{j
\mu}+{{{\cal T}_{r}^{{}^{{}^{^{}{\!\!\!\!\scriptstyle{em}}}}}}}^{j
\mu}\right)\right]=0.\]
\newpage
\section{Asymptotically AdS black holes}\label{S3}
We apply the  field equations of extended  teleparallel gravity $f(T)$, Eq. (\ref{q8}), to the flat $D$-dimensional  spacetime horizon, which directly gives rise to the vielbein  written in cylindrical  coordinate ($t$, $r$, $\phi_1$, $\phi_2$,$\cdots$ $\phi_{D-2}$) as follows \cite{CGSV13}:
\begin{equation}\label{tetrad}
\hspace{-0.3cm}\begin{tabular}{l}
  $\left({e_{i}}^{\mu}\right)=\left( \sqrt{N(r)}, \; \frac{1}{\sqrt{N_1(r)}}, \; r, \; r, \; r\;\cdots \right)$
\end{tabular}
\end{equation}
where $N(r)$ and $N_1(r)$ are two unknown functions of $r$. Substituting from Eq. (\ref{tetrad}) into Eq. (\ref{Tor_sc}), we evaluate the torsion scalar as\footnote{For abbreviation we will write $N(r)\equiv N$,  \ \ $N_1(r)\equiv N_1$, \ \ $N'\equiv\frac{dN}{dr}$ and $N'_1\equiv\frac{dN_1}{dr}$ .}
\begin{equation}\label{df}
T=2(D-2)\frac{N'N_1}{rN}+(D-2)(D-3)\frac{N_1}{r^2}.
\end{equation}
Applying Eq. (\ref{tetrad}) to the field equation (\ref{q8}) when ${{{\cal
T}^{{}^{{}^{^{}{\!\!\!\!\scriptstyle{em}}}}}}}^\nu_\mu=0$ we get the following non-vanishing components:
\begin{eqnarray}\label{df1}
& & H^r{}_r= 2Tf_T+2\Lambda-f=0,\nonumber\\
& &  H^{\phi_1}{}_{\phi_1}= H^{\phi_2}{}_{\phi_2}=\cdots \cdots =H^{\phi_{D-2}}{}_{\phi_{D-2}}=   \frac{f_{TT} [r^2T+(D-2)(D-3)N_1]T'}{(D-2)r}+\frac{f_T}{2r^2{N}^2}\Biggl\{2r^2NN_1N''\nonumber\\
& &-r^2N_1N'^2+2(2D-5)rNN_1N' +r^2NN'N'_1+2(D-3)N^2[2(D-3)N_1+rN'_1]\Biggr\}-f+2\Lambda=0, \nonumber\\
& & H^t{}_t=\frac{2(D-2)N_1f_{TT} T'}{r}+\frac{(D-2)f_T}{r^2N}\Biggl\{2(D-3)NN_1+rN_1N'+rNN'_1\Biggr\}-f+2\Lambda=0.\nonumber\\
& &
\end{eqnarray}
Now we are going to find a general solution to the above differential equations using a specific form of $f(T)$, i.e., $f(T)=T+\beta T^2$.
For this specific form of $f(T)$,  Eqs. (\ref{df1}) take the form
\begin{eqnarray}
& & H^r{}_r=  T+3\beta T^2+2\Lambda=0,\nonumber\\
& & H^{\phi_1}{}_{\phi_1}= H^{\phi_2}{}_{\phi_2}=\cdots \cdots H^{\phi_{D-2}}{}_{\phi_{D-2}}=  \frac{2\beta[r^2T+(D-2)(D-3)N_1]T'}{(D-2)r}+\frac{(1+2\beta T)}{2r^2{N}^2}\Biggl\{2r^2NN_1N''\nonumber\\
& &-r^2N_1N'^2+2(2D-5)rNN_1N'+r^2NN'N'_1+2(D-3)N^2[2(D-3)N_1+rN'_1]\Biggr\} -T-\beta T^2+2\Lambda=0,\nonumber\\
& &H^t{}_t= \frac{4(D-2)\beta N_1T'}{r}+\frac{(1+2\beta T)(D-2)}{r^2N}\Biggl\{2(D-3)NN_1+rN_1N'+rNN'_1\Biggr\}-T-\beta T^2+2\Lambda=0.\nonumber\\
&&
\end{eqnarray}
A general $D$-dimension  solution  of Eq. (18) is

\begin{eqnarray}
& &  N(r)=-\frac{r^2}{6(D-1)(D-2)\beta}-\frac{m}{r^{D-3}}, \qquad \quad
N_1(r)=\frac{1}{N(r)},  \nonumber\\
\end{eqnarray}
where $m$ is the mass parameter and we choose $\Lambda=\frac{1}{24\beta} $ to get a unique solution\footnote{The cosmological constant for these solutions has two values $ \frac{-1 \pm \sqrt{1-24\alpha \, \Lambda}}{12 \alpha}$}.

\section{A new charged AdS black hole solution}\label{S9}

Using the {\it D} dimensional spacetime of Eq. (\ref{tetrad}) with a vector potential $A = V(r) dt$, the field equations have the following non-vanishing components:
\begin{eqnarray}
& & H^r{}_r= 2Tf_T+2\Lambda-f+\frac{2V'^2(r)N_1}{N}=0,\nonumber\\
& & H^{\phi_1}{}_{\phi_1}= H^{\phi_2}{}_{\phi_2}=\cdots \cdots =H^{\phi_{D-2}}{}_{\phi_{D-2}}=  \frac{f_{TT} [r^2T+(D-2)(D-3)N_1]T'}{(D-2)r}+\frac{f_T}{2r^2{N}^2}\Biggl\{2r^2NN_1N''\nonumber\\
& & -r^2N_1N'^2+4(D-3)^2N^2N_1+2(2D-5)rNN_1N'+r^2NN'N'_1+2(D-3)rN^2N'_1\Biggr\}\nonumber\\
& & -f+2\Lambda-\frac{2V'^2(r)N_1}{N}=0, \nonumber\\
& & H^t{}_t= \frac{2(D-2)N_1f_{TT} T'}{r}+\frac{(D-2)f_T[2(D-3)NN_1+rN_1N'+rNN'_1]}{r^2N}-f+2\Lambda+\frac{2V'^2(r)N_1}{N}=0,\nonumber\\
& &
\end{eqnarray}
where $V'=\frac{dV}{dr}$ and as before we set $\Lambda=\frac{1}{24\beta} $.
The general  {\it D}-dimensional solution of the above differential equations takes the form
\begin{eqnarray}
& &  N(r)=\frac{r^2(D-3)^4c_2{}^4}{(D-1)(D-2)(2D-5)^2c_3{}^2}+\frac{c_1}{r^{D-3}}+\frac{3(D-3)c_2{}^2}{(D-2)r^{2(D-3)}}+\frac{2(D-3)c_2c_3}{(D-2)r^{3D-8}},\nonumber\\
& &
 N_1(r)=\frac{1}{f(r)N(r)}, \qquad  \textrm {where} \qquad f(r)=-\frac{(2D-5)^2c_3{}^2\left[1+\frac{(2D-5)c_3}{c_2(D-3)r^{D-2}}\right]^2}{6\beta (D-3)^4c_2{}^4},\nonumber\\
 & &  V(r)=\frac{c_2}{r^{D-3}}+\frac{c_3}{r^{2D-5}}.\end{eqnarray}
To get an asymptotically AdS or dS solution we have to set \begin{eqnarray} & & c_3{}^2=\frac{-6(D-3)^4c_2{}^4 \beta}{(2D-5)^2},\end{eqnarray}
otherwise the solution  have no clear asymptotic behavior. As a result, the monopole momentum is related to the quadrupole momentum of the solution. In this case, one gets
\begin{eqnarray} & &N(r)=\frac{r^2}{6(D-1)(D-2)\left|\beta\right|}-\frac{m}{r^{D-3}}+\frac{3(D-3)q{}^2}{(D-2)r^{2(D-3)}}+\frac{2\sqrt{6\left|\beta\right|}(D-3)^3q{}^3}{(2D-5)(D-2)r^{3D-8}},\nonumber\\
& &
 N_1(r)= \frac{1}{f(r)N(r)}, \qquad f(r)=\left[1+\frac{(D-3)q\sqrt{6\left|\beta\right|}}{r^{D-2}}\right]^2,\qquad V(r)=\frac{q}{r^{D-3}}+\frac{(D-3)^2q{}^2\sqrt{6\left|\beta\right|}}{(2D-5)r^{2D-5}},\nonumber\\
& &\end{eqnarray}
 where we set $c_1=-m$, and $c_2=q$, which is the monopole momentum. The quadrupole moment is $Q= \frac{(D-3)^2q{}^2\sqrt{6 \left| \beta \right|}}{(2D-5)}$.  As one can notice Eq. (22) tells us that $\beta$  have a negative value otherwise we get an unphysical solution.\\

It is important here to comment on the charged black solutions obtained in \cite{CGSV13}. Note that differential equations (20) are different from those derived in \cite{CGSV13} from many aspects:\vspace{0.2cm}\\
i) The disappearance of the term $f(T)$ from Eqs. (5$\cdot$13) and (5$\cdot$14).\vspace{0.2cm}\\
ii) The terms of the charges in Eqs.  (5$\cdot$12), (5$\cdot$13) and (5$\cdot$14) are different from the present Eqs. (20).

 Furthermore, the solution (23) generalizes the solution in \cite{CGSV13}.  As it is clear from Eq. (23),  the potential $V(r)$  depends on a monopole and quadrupole momenta and by setting $q=0$ both momenta  vanish and we get a non-charged solution. On the other hand,  in \cite{CGSV13}, the charged solution  depends only on the monopole.

\section{The main features of the solution}\label{S5}
Let us now discuss  some relevant features of the charged solution presented in  the previous Section.\\

The metric of the vielbein  (23) takes the form

\ba \A \A ds{}^2=\Biggl[r^2\Lambda_{ef}-\frac{m}{r^{D-3}}+\frac{3(D-3)q^2}{(D-2)r^{2(D-3)}}+\frac{2\sqrt{6\left|\beta\right|}(D-3)^3q^3}{(2D-5)(D-2)r^{3D-8}}
\Biggr]dt^2\nonumber\\
-\A\A\frac{dr^2}{\left[1+\frac{(D-3)q\sqrt{6\left|\beta\right|}}{r^{D-2}}\right]^2\Biggr[r^2\Lambda_{ef}-\frac{m}{r^{D-3}}+\frac{3(D-3)q^2}{(D-2)r^{2(D-3)}}
+\frac{2\sqrt{6\left|\beta\right|}(D-3)^3q^3}{(2D-5)(D-2)r^{3D-8}}\Biggr]} -r^2\sum_{i=1}^{D-2}d\phi^2_i,\nonumber\\
\A\A\ea
where $\Lambda_{ef}= \frac{1}{6(D-1)(D-2)\left|\beta\right|}$. Eq. (23) shows clearly that the metric of the charged solution is asymptotically  AdS. Notice that there is no corresponding TEGR solution upon taking the limit $\beta \rightarrow 0$, which means this charged solution has no analogue in GR or TEGR. By taking the limit $q \rightarrow 0$, we get the AdS non-charged black holes presented in section (\ref{S3}). Notice that although these asymptotically AdS charged solutions have different $g_{tt}$ and $g^{rr}$ component for the metric, they have coinciding Killing  and event horizons.

\underline{Singularity:}\vspace{0.2cm}\\
 Here we derive physical singularities by calculating curvature and torsion invariants. Since the function $f(r)$ could have roots (when $q<0$), which we call $r_n$, one has to consider  the behavior of curvature invariants close to these roots. By calculating the Kretschmann scalar as function of the radial coordinate $r$, we found that the scalar is well behaved. Now calculating the various curvature and torsion invariants, one obtains

 \begin{eqnarray} \A \A R^{\mu \nu \lambda \rho}R_{\mu \nu \lambda \rho}= F_1(r)\,\left(\frac{1}{r^{2(D-2)}}\right), \qquad  R^{\mu \nu}R_{\mu \nu}=  F_2(r)\,\left(\frac{1}{r^{2(D-2)}}\right), \nonu
  \A \A R=  F_3(r)\,\left(\frac{1}{r^{(D-2)}}\right), \qquad  T^{\mu \nu \lambda}T_{\mu \nu \lambda} = F_4(r)\, \left(\frac{1}{r^{(D-2)}}\right), \qquad T^\mu T_\mu = F_5(r)\, \left(\frac{1}{r^{(D-2)}}\right),
 \nonu
\A \A T(r)=F_6(r)\, \left(\frac{1}{r^{(D-2)}}\right),\end{eqnarray}
where $R^{\mu \nu \lambda \rho}R_{\mu \nu \lambda \rho}$, $R^{\mu \nu}R_{\mu \nu }$, $R$, $T^{\mu \nu \lambda}T_{\mu \nu \lambda}$ $T^{\mu }T_{\mu }$ and $T$ are the Kretschmann scalar,  the Ricci tensor square, the Ricci scalar, the torsion tensor square,  the torsion square vector and the torsion scalar:  $F_i(r)$ are polynomial functions in $r$.
 The above invariants show that:\vspace{0.1cm}\\
 a) There is a singularity at $r=0$ which is a curvature singularity. \vspace{0.1cm}\\
 b) In the charged case,  the torsion scalar has the form \be T=\frac{r^{(D-2)}-2q^2(D-3){\sqrt{6\left|\beta\right|}}}{6\left|\beta\right| r^{(D-2)}},\ee which shows that the scalar torsion has singularity at $r=0$. Close to $r=0$,  the behavior of the Kretschmann scalar, the Ricci tensor square and the Ricci scalar for the charged solution  is given by $K=R_{\mu \nu}R^{\mu \nu} \sim r^{-2(D-2)}$, $R=T^{\alpha \beta \gamma}T_{\alpha \beta \gamma}=T^{\alpha}T_{\alpha}=T\sim  r^{-(D-2)}$ in contrast with the  solutions of the  Einstein-Maxwell theory in both GR and TEGR which have $K =R_{\mu \nu}R^{\mu \nu}\sim r^{-2D}$ and $R=T^{\alpha \beta \gamma}T_{\alpha \beta \gamma}=T^{\alpha}T_{\alpha}=T \sim  r^{-D}$. This shows clearly that the singularity is much milder than the one obtained in GR and TEGR for the charged case. This result raises the question if these singularities are {\it weak singularities},  according to Tipler and Krolak \cite{tipler,krolak},  and if  it is possible to extend geodesics beyond these regions. This topic will be discussed in forthcoming studies.

\underline{Energy:}\vspace{0.2cm}\\
  Let us now calculate the energy related to  the  charged black holes given by Eqs. (23). Using Eq. (14),  it is possible to derive  the  components of energy  in the  solution (19).   We get:
\be S^{001}=\frac{(D-2)}{2r}.  \ee
 Substituting Eq. (27) into Eq. (14),  we get the energy in the form
\be P^0=E= \frac{(D-2)\Omega_{D-2}[m-\Lambda_{ef}\,r^{(D-1)}]}{2\kappa }
=\frac{(D-2)[m-\Lambda_{ef}\,r^{(D-1)}]}{4(D-3)G_D}, \ee
where the value of $\kappa$ has been used in the second equation of Eq. (28). The value of energy of Eq. (28) is  therefore divergent, so  we have to use a regularization  procedure  to obtain a finite value. The regularized expression of Eq. (14) takes the form
\be P^a:=\frac{1}{\kappa}\int_V d^{D-2}x  \left[e{S}^{a 0
0} f_T\right]-\frac{1}{\kappa}\int_V d^{D-2}x \left[e{S}^{a 0
0} f_T\right]_{AdS},\ee where {\it AdS} means calculated for pure Anti-de-Sitter space.  Using (30) in solution (19),  we get
\be E =\frac{(D-2)\Omega_{D-2}m}{2\kappa }= \frac{(D-2)m}{4(D-3)G_D},\ee
which is a finite value and clearly shows  that the energy is depending on  the coefficient of the higher order torsion terms.  For  the charged solutions given in  (21), and by using the same procedure adopted  for the non-charged case,  we get
 \ba \A \A E=\frac{(D-2)m}{4(D-3)G_D}-\frac{(D-3)q^2 }{2G_D r^{D-3}}+\frac{(D-3)^3\sqrt{6\left|\beta\right|}q^3 }{3(2D-5)G_D r^{2D-5}}+O\Biggl(\frac{1}{r^3}\Biggr).\ea
This shows the contributions of monopole and quadrupole potential energies to the total energy at  large distances $r$. Notice  only term  depending on $\beta$ is the quadrupole term. Considering the limit $r\rightarrow \infty$, we get the total energy measured  by a stationary observer at infinity.
\section{The first law of thermodynamics}\label{S11}

There is a great deal of work in analyzing the behavior of the horizon thermodynamics in modified  theories of gravity. In a wide category of these theories, one gets solutions with horizons and can connect the temperature and entropy with the surface gravity and the area of the outer horizon. Let us now check the validity of the first law of thermodynamics in $f(T)$ gravity using the  charged solution derived above.

To investigate the violation of the first law of thermodynamics of the black hole (23), let us follow the analysis performed by Miao et al. \cite{MLM}. In this work the authors use the  Jacobson thermodynamics approach \cite{jacobson}, which has been generalized in \cite{akbar,adel+ahmed},  to formulate the first law through rewriting the non-symmetric field equations (9) into symmetric and skew symmetric parts as
\begin{eqnarray}\label{q88}
\A \A L_{(\mu\nu)}:=S_{\mu \nu \rho} \partial^{\rho} T f_{TT}+f_T \left[G_{\mu \nu} -\frac{1}{2}g_{\mu \nu}T\right]
+\frac{f-2\Lambda}{2}g_{\nu \mu} =\frac{\kappa {{{\cal
T}^{{}^{{}^{^{}{\!\!\!\!\scriptstyle{em}}}}}}}^\nu_\mu}{2},\nonu
\A \A L_{[\mu \nu]}:=S_{[\mu \nu] \rho} \partial^{\rho} T f_{TT}=0.\end{eqnarray}
Assuming an exact Killing vector, they have shown that for a heat flux $\delta Q$  passing through  the black hole horizon, it  is
\be  \delta Q=\frac{\kappa}{2\pi}\left[\frac{f_T dA}{4}\right]^{d\lambda}_0+\frac{1}{\kappa}\int_H k^\nu  f_{TT} \ T_{,\mu}(\xi^\rho S_{\rho \nu}{}^ \mu-\nabla_\nu \xi^\mu),\ee where $H$ stands for the black hole horizon which is equal to $(D - 2)$-dimensional boundary of the hypersurface at infinity.
The authors have shown that the first term in Eq. (33) can be rewritten as $T\delta S$ \cite{ MLM}. Thus, if the second term in Eq. (33) is not vanishing then there will be a violation of the first law of thermodynamics. Miao et al.  \cite{MLM} have explained  that the second term in Eq. (34) cannot be equal to zero. Therefore, if we want to satisfy the first law of thermodynamics, we must have either  $f_{TT}=0$ which gives the TEGR (GR) theory, or $T=constant$. Indeed, AdS black hole solution (19) satisfies the fact that $T=constant$ and therefore, the first law of thermodynamics of this black hole is satisfied. However, the new charged solution (23) enforces the torsion scalar to be non-trivial, i.e., not constant. Therefore, according to Eq. (33),  solution (23) violates the first law of thermodynamics.  Explicitly calculating the violation term in (33),  it is not vanishing and proportional to the electric charge $q$. Due to this feature,   the first law of thermodynamics is violated.

 \section{Conclusions}\label{S12}
In this work, we present a new charged solution in Maxwell-$f(T)$ gravitational theory for any dimension $D\geq 4$. The exact solution is achieved for  $f(T)=T+\beta T^2$, where $\beta <0$ and possesses  some interesting features. First of all, the solution has a monopole and quadrupole term which are related by requiring the metric is asymptotically AdS. This fact generalizes the result presented in \cite{CGSV13} where   only the  monopole term is present.
Secondly, we have studied the singularity of this black hole and have shown that all the invariants constructed from the curvature and torsion have a singularity at $r=0$. This singularity is milder than the one of a charged black hole in GR and TEGR. The asymptotic behavior of the Kretschmann invariant and the Ricci tensor squared, and the Ricci scalar  have the form $K=R_{\mu \nu}R^{\mu \nu} \sim r^{-2(D-2)}$, $R\sim  r^{-(D-2)}$  in contrast with, the known solutions in Einstein-Maxwell theory in both GR and TEGR. Also the non-charged solution derived in this study, Eq. (19), behaves as $K =R_{\mu \nu}R^{\mu \nu}\sim r^{-2D}$ and $R \sim  r^{-D}$. Moreover, in spite that the charged black hole has different components of $g_{tt}$ and $g^{rr}$, both have a coinciding Killing and event horizons. We have calculated the total energy of the charged solution using the generalization of the energy-momentum tensor and have shown that the resulting form  depends on the mass of the black hole. Finally, we have shown that the charged  black hole violates the first law of thermodynamics according  to the discussion given in \cite{MLM}.
From a genuine physical point of view, these kind of objects can contribute in the debate to establish on what the most reliable representation of gravity is, i.e.  the  curvature or torsion picture. As discussed in \cite{CCMS}, being GR and TEGR substantially equivalent, the ground of debate should be shifted to $f(R)$ and $f(T)$ models since these two theories are substantially inequivalent. As shown in \cite{diego}, also fundamental structures like gravitational waves are substantially different in $f(R)$ and in $f(T)$ formulations. A deep understanding of black hole features could be of great interest to solve the debate.
\subsection*{Acknowledgments}
We would like to thank  E. N. Saridakis for useful discussions on the topic. S. Capozziello acknowledges the COST Action CA15117 (CANTATA).
 This work is partially supported by the Egyptian Ministry of Scientific Research under project No. 24-2-12.
\newpage

\end{document}